\documentclass[10pt,a4paper]{article}

\begin{document}

\title{Born's group and Generalized isometries\\
{\normalsize Edited version of Ll. Bel (1994)}}
\author{Ll. Bel\thanks{e-mail:  wtpbedil@lg.ehu.es}}

\maketitle

\begin{abstract}
We define the Born group as the group of transformations that leave
invariant the line element of Minkowski's spacetime written in terms of
Fermi coordinates of a Born congruence. This group depends on three
arbitrary functions of a single argument. We construct implicitly the finite
transformations of this group and explicitly the corresponding infinitesimal
ones. Our analysis of this group brings out the new concept of Generalized
group of isometries. The limitting cases of such groups being, at one end,
the Groups of isometries of a spacetime metric and, at the other end, the
Group of diffeomorphisms of any spacetime manifold. We mention two examples
of potentially interesting generalizations of the Born congruences.
\end{abstract}

\section{Fermi, Born and Killing congruences}

{\em Fermi-Walker transport}, (\cite{fermi}, \cite{walker}, \cite{mast}). Let $L$
be a worldline and $\overline{u}^\alpha $ be its tangent unit vector field.
Let $\overline{e}^\alpha $ be a vector field defined along $L$. By
definition this vector field is Fermi-Walker transported if it is a solution
of the differential equation:

\begin{equation}
\label{transport}{\frac{D\overline{e}^\alpha }{{d\tau }}}\equiv {\frac{%
\nabla \overline{e}^\alpha }{{d\tau }}}+(\overline{b}^\alpha \overline{u}%
_\rho -\overline{u}^\alpha \overline{b}_\rho +\frac 12\overline{\Omega }%
_{.\rho }^\alpha )e^\rho =0,
\end{equation}
where $\overline{\Omega }_{\alpha \rho }$ is a skewsymmetric 2-rank tensor
orthogonal to $\overline{u}^\alpha $:%
$$
\overline{\Omega }_{\alpha \rho }=-\overline{\Omega }_{\rho \alpha },\quad
\overline{\Omega }_{\alpha \rho }\overline{u}^\alpha =0
$$
defined along $L$, and where:

$$
{\frac{\nabla \overline{e}^\alpha }{{d\tau }}}\equiv {\frac{d\overline{e}%
^\alpha }{{d\tau }}}+\overline{\Gamma }_{\beta \gamma }^\alpha \overline{u}%
^\beta \overline{e}^\gamma ,\qquad \overline{b}^\alpha \equiv {\frac{\nabla
\overline{u}^\alpha }{{d\tau }}=\frac{d\overline{u}^\alpha }{{d\tau }}}+%
\overline{\Gamma }_{\beta \gamma }^\alpha \overline{u}^\beta \overline{u}%
^\gamma ,
$$
$\tau $ being the proper time measured along $L$ with arbitrary origin, and $%
\overline{\Gamma }_{\beta \gamma }^\alpha $ being the restriction of the
Christoffel symbols to $L$.

It follows from this definition that the tangent unit vector field $%
\overline{u}^\alpha $ is a solution of Eq.(\ref{transport}). Therefore it can
be thought of as being Fermi-Walker propagated along $L$. Another important
property is that the scalar product of any two Fermi-Walker propagated
vector fields is constant along $L$.

{\em Fermi coordinates}, (\cite{fermi},\cite{manasse},\cite{Ni}). Let us
consider a worldline $L$ and let $\overline{e}_{0i}^\alpha $ be three
vectors defined at some event $O$ on $L$ such that:

$$
\overline{e}_{0i}^\alpha \overline{e}_{0j\alpha }=\delta _{ij},\quad
\overline{u}_0^\alpha \overline{e}_{i\alpha }=0,
$$
$\overline{u}_0^\alpha $ being the unit vector tangent to $L$ at $O$.
Fermi-Walker propagating each of these three spacelike vectors along $L$ we
obtain at each event of $L$ an orthonormal frame of reference adapted to $L$:

$$
\overline{e}_i^\alpha \overline{e}_{j\alpha }=\delta _{ij},\quad \overline{u}%
^\alpha \overline{e}_{i\alpha }=0.
$$
We shall say that $T$ is the Normal tube of $L$ if for each event $E$ of $T$
there exists one and only one spacelike geodesic passing through $E$ and
being orthogonal to $L$. We shall note this geodesic $G(E_0,E)$, $E_0$ being
its intersection with $L$.

By definition the Fermi coordinates with Baseline $L$ of any event $E$ of $T$
are:%
$$
z^0=\tau ,\quad z^i=n^\alpha e_\alpha ^ir
$$
where $\tau $ is the proper-time interval measured along $L$ from $O$ to $%
E_0 $, $n^\alpha $ is the unit tangent vector to $G(E_0,E)$ at $E_0$, and $r$
is the geodesic distance from $E_0$ to $E$.

{\em Fermi congruences}. Given a worldline $L$, a Fermi congruence $C(L)$
with Baseline $L$ is the congruence with parametric equations:%
$$
z^0=\tau ,\quad z^i=c^i
$$
$z^\alpha $ being a system of Fermi coordinates with Baseline $L$ and $c^i$
three arbitrary constants.

Equivalently a Fermi congruence $C(L)$ with Baseline $L$ can be defined as a
congruence such that the orthogonal geodesic distance $G(E_0,E)$ remains
constant when the event $E$ is displaced along the worldline of the
congruence passing through $E$

Let $C$ be a timelike congruence defined on some world tube $T$, and $%
u^\alpha $ be its unit tangent vector. There are three basic geometrical
objects that can be associated to it besides its {\em Projector}:

\begin{equation}
\label{projector}\hat g_{\alpha \beta }=g_{\alpha \beta }+u_\alpha u_\beta .
\end{equation}

The {\em Curvature}:

\begin{equation}
b^\alpha =u^\rho \nabla _\rho u^\alpha ,\quad u^\alpha b_\alpha =0.
\end{equation}

The {\em Deformation rate} of the congruence, which is the symmetric 2-rank
tensor orthogonal to $u^\alpha $:

\begin{equation}
\label{deformation}\Sigma _{\alpha \beta }=\hat \nabla _\alpha u_\beta +\hat
\nabla _\beta u_\alpha ,\quad \Sigma _{\alpha \beta }u^\alpha =0,
\end{equation}
where:
$$
\hat \nabla _\alpha u_\beta \equiv \hat g_\alpha ^\rho \hat g_\beta ^\sigma
\nabla _\rho u_\sigma.
$$

And the {\em Rotation rate} of the congruence, which is the skewsymmetric
2-rank tensor orthogonal to $u^\alpha $:

\begin{equation}
\label{rotation}\Omega _{\alpha \beta }=\hat \nabla _\alpha u_\beta -\hat
\nabla _\beta u_\alpha ,\quad \Omega _{\alpha \beta }u^\alpha =0.
\end{equation}

{\em Chorodesics. }We shall say that a curve $x^\alpha =x^\alpha (\lambda )$
is a chorodesic of a timelike congruence $C$ if it is a solution of the
following system of differential equations:

\begin{equation}
\label{eqschoro}\frac{dx^\alpha }{d\lambda }=p^\alpha \quad \frac{dp^\alpha
}{d\lambda }+\Gamma _{\lambda \mu }^\alpha p^\lambda p^\mu =\frac 12u^\alpha
\Sigma _{\lambda \mu }p^\lambda p^\mu ,
\end{equation}
where $u^\alpha $ is the tangent vector to $C$ and $\Sigma _{\lambda \mu
}$ is its Deformation rate. If $\Sigma _{\lambda \mu }=0$ then the
chorodesics of $C$ and the geodesics of the spacetime coincide.

{\bf Lemma 1}
\label{chorodesics}: If a spacelike chorodesic is orthogonal to a worldline of
a congruence $C$ then it is orthogonal to all the worldlines of the
congruence $C$ that it crosses.

In fact from Eqs.(\ref{eqschoro}) it follows that:

$$
\frac d{d\lambda }(u_\alpha p^\alpha )=p^\alpha p^\beta \left(\nabla _\alpha
u_\beta -\frac 12\Sigma _{\alpha \beta }\right);
$$
or, using the definition (\ref{deformation}) of $\Sigma _{\alpha \beta }$:

$$
\frac d{d\lambda }(u_\alpha p^\alpha )=(u_\alpha p^\alpha )(b_\beta p^\beta
).
$$
Therefore if $u_\alpha p^\alpha $ is zero at one event it will remain zero
all along the chorodesic passing through this event and having $p^\alpha $
as tangent.

{\bf Lemma 2}\label{fermi2born}: If $C$ is a Fermi congruence with Baseline $L$ then the
restriction of the Deformation rate tensor on $L$ is zero: $\overline{\Sigma
}_{\alpha \beta }=0$.

We shall say that $C$ is an {\em Homogeneous Fermi congruence} on $T$ if
each worldline of $C$ can be viewed as a Baseline of $C.$

{\em Born congruences}, (\cite{born},\cite{synge2},\cite{synge},\cite{belcoll}).
Born congruences were initially defined as those congruences for which the
infinitesimal orthogonal distance between neighboring worldlines was
constant along the worldlines. It follows from this definition that the
deformation rate tensor of a Born Congruence is zero:

\begin{equation}
\label{borneqs}\Sigma _{\alpha \beta }=0.
\end{equation}

{\bf Lemma 3}
\label{bornandgeodesics}: If a spacelike geodesic is orthogonal to a worldline
of a Born congruence then it is orthogonal to all the worldlines of the
congruence that it crosses. Reciprocally, if $C$ is a congruence such that
if a geodesic is orthogonal to a worldline of $C$ then it is orthogonal to
all the worldlines of $C$ that it crosses, then $C$ is a Born congruence.

The first part of this Lemma is a corollary of lemma \ref{chorodesics}.

{\bf Theorem 1}
\label{finiteborn}: If $L_0$ and $L_1$ are two worldlines of a Born congruence
then the length of any geodesic orthogonal arc $G(E_0,E_1)$ intercepted by
these two worldines remains constant when $E_0$ and $E_1$ move
accordingly along the two worldlines.

This follows immediately from lemma \ref{bornandgeodesics}.

{\bf Theorem 2}:
Every Homogeneous Fermi congruence is a Born congruence, and reciprocally
every Born congruence is an Homogeneous Fermi congruence.

The first part of this Theorem follows immediately from the Lemma \ref
{fermi2born}. The second part can be proved as follows: let $L_0$ and $L_1$%
be two worldlines of a Born congruence. From the Lemma \ref{bornandgeodesics}
it follows that the family of geodesics which intersect both worldlines and
are orthogonal to $L_0$ are also orthogonal to $L_1$. The length of any
geodesic arc $G(E_0,E_1)$ intercepted by these two worldines will be given
by the integral:

$$
L=\int_{E_0}^{E_1}\sqrt{g_{\alpha \beta }dx^\alpha dx^\beta },
$$
calculated along $G(E_0,E_1)$. The variation of this integral when $E_0$ and
$E_1$ move along $L_0$ and $L_1$is :

$$
\delta L=g_{\alpha \beta }(x_1^\rho )\delta x_1^\alpha n_1^\beta -g_{\alpha
\beta }(x_0^\rho )\delta x_0^\alpha n_0^\beta
$$
$n_0^\alpha $ and $n_1^\alpha $ being the unit tangent vectors of $%
G(E_0,E_1) $ at the events $E_0$ and $E_1$. Since at both ends $\delta
x^\alpha $ and $n^\alpha $ are orthogonal, it follows that $\delta L=0$
which proves that the geodesic arc length of $G(E_0,E_1)$ remains constant.
Since $L_0$ and $L_1$ are arbitrary it follows that the congruence is an
homogeneous Fermi congruence.

{\em Killing congruences}, (\cite{killing}). A Killing vector field $\xi
^\alpha $ is a generator of an infinitesimal symmetry of spacetime. It
satisfies the Killing equations:

$$
\nabla _\alpha \xi _\beta +\nabla _\beta \xi _\alpha =0
$$
A Killing congruence $K$ is a congruence tangent to a Killing vector field.
If $K$ is timelike and $u^\alpha $ is the corresponding unit tangent vector
field to $K$ then it can be characterized by the following equations:

$$
\Sigma _{\alpha \beta }=0,\quad \partial _\alpha b_\beta -\partial _\beta
b_\alpha =0
$$
It follows from this last characterization that any Killing congruence is a
Born congruence. We shall call Pure Born those Born congruences which are
not Killing congruences, i.e., those for which $b_\alpha $ is not a gradient

{\bf Theorem 3},
(\cite{herglotz},\cite{noether}): In Minkowski spacetime, $M_4,$ any Born
congruence $C$ is either a Killing congruence or an irrotational congruence.
More precisely: if it is a Killing congruence then the tangent unit vector
field is colineal to one of the generators of a Poincar\'e transformation.
If it is a Pure Born congruence then it is the irrotational Fermi congruence
with Baseline any wordline of the congruence.

From the last part of this theorem it follows that if $x^\alpha $ is a
Galilean system of coordinates of $M_4$:

\begin{equation}
\label{Minkowski}ds^2=-dt^2+\delta _{ij}dx^idx^j,
\end{equation}
and $\bar y^\alpha (\tau )$ are the parametric equations of a timelike
worldline $L$ then, using Fermi coordinates $(\tau ,z^i)$ with Baseline $L$,
the parametric equations of the unique Pure Born congruence containing $L$
are:

\begin{equation}
\label{fermiparametric}x^\alpha =\bar e_i^\alpha (\tau )z^i+\bar y^\alpha
(\tau ),
\end{equation}
where $\bar e_i^\alpha (\tau )$ is a system of orthonormal vectors
Fermi-Walker propagated along $L$.

Differentiating these equations and substituting in (\ref{Minkowski}) the line
element of $M_4$ becomes\cite{moller}:

\begin{equation}
\label{MinkowskiFermi}ds^2=-[1+a_i(\tau )z^i]^2d\tau ^2+\delta
_{ij}dz^idz^j,
\end{equation}
where $a_i=\bar e_i^\alpha \bar b_\alpha $.

\section{Poincar\'e and Born's group\label{borngroup}}

Let us consider Minkowski's spacetime $M_4$, a system of Galilean coordinates,
and the Poincar\'e group:

$$
x^{\prime \alpha }=L^\alpha (x^\beta ,\Lambda _I)\equiv L_\beta ^\alpha
(x^\beta -A^\beta )
$$
where $L_\beta ^\alpha $ are Lorentz matrices. The line element (\ref
{Minkowski}) is invariant under this group which means that in terms of the
new Galilean coordinates $x^{\prime \alpha }$ the same line element becomes:

\begin{equation}
\label{minkowski2}ds^2=-dt^{\prime 2}+\delta _{ij}dx^{\prime i}dx^{\prime j}.
\end{equation}

Let us assume that we perform the coordinate transformation (\ref
{fermiparametric}), which we write for short as:

$$
z^\alpha =F^\alpha (x^\rho ),\qquad x^\rho =\tilde F^\rho (z^\alpha ),
$$
leading to Minkowski's line element (\ref{MinkowskiFermi}), and let us
consider the family of composite transformations:
$$
z^{\prime\alpha }=F^\alpha \{L^\beta [\tilde F^\gamma (z^\delta )]\}
$$
By construction this family of transformations is still a group. More
precisely it is the realization of the Poincar\'e group which leaves
invariant the line element of Minkowski spacetime written as in (\ref
{MinkowskiFermi}). This meaning that in terms of the new coordinates we shall
have:

\begin{equation}
\label{MinkowskiFermi2}ds^2=-[1+a_k(\tau ^{\prime })z^{\prime k}]^2d\tau ^{\prime 2}+\delta _{ij}dz^{\prime i}dz^{\prime j}
\end{equation}
Notice that the functions $a_k$ are the same in expressions (\ref
{MinkowskiFermi}) and (\ref{MinkowskiFermi2}).

We define the Fermi realization of the Born group as the family of
transformations:

$$
z^{\prime \alpha }=B^{\prime \alpha }(z^\rho ),\qquad z^\rho =B^\rho
(z^{\prime \alpha })
$$
which leave invariant the form of the line element (\ref{MinkowskiFermi}):
$$
ds^2=-[1+a_k^\prime (\tau ^\prime )z^{\prime k}]^2d\tau
^{\prime 2}+\delta _{ij}dz^{\prime i}dz^{\prime j}
$$
without necessarily leaving invariant the functions $a_k$. That is to say:
in general $a_k^\prime \not=a_k$.

By construction the Born group contains as subgroup the Poincar\'e group. It
contains also an isotropy subgroup which leaves invariant the Born
congruence without leaving invariant the Baseline on which the Fermi
coordinates are based. This subgroup which depends on three parameters $%
\lambda ^i$ is:

$$
z^{\prime i}=z^i-\lambda ^i\qquad \tau^\prime =\tau +v_k(\tau
)\lambda ^k
$$
where:
$$
v_k(\tau )=\int_0^\tau a_k(u)du
$$
Under a transformation of this subgroup the functions $a_k$ become:
$$
a_k^\prime (u,\lambda ^i)={\frac{a_k(u)}{{1+a_j(u)\lambda ^j}}}
$$
This subgroup is abelian and translates the Baseline from a worldline of the
congruence to another. It does not have to be confused with the translation
subgroup of the Poincar\'e subgroup which leaves invariant the functions $%
a_k $.

{\em Infinitesimal transformations}. Let $L$ and $L^{\prime }$ be two
neigbouring worldlines in Minkowski's spacetime

$$
\bar y^{\prime \alpha }(\tau ^\prime)=\bar y^\alpha (\tau^\prime)
+\delta \bar y^\alpha (\tau ^\prime)
$$
Let us assume that the origins of the proper times are close to each other
and that at these origins two adapted orthonormal frames of reference have
been defined and are such that:
$$
{\bar e}_i^{\prime \alpha }(0)=\bar e_i^\alpha (0)+\delta \Omega
_i^j(0)e_j^\alpha (0)
$$
Let us finally assume that the curvatures of both worldlines do not differ
much when considered at the same value of their arguments:

$$
a^\prime_i=a_i+\zeta_i
$$
If $(\tau ,z^i)$ and $(\tau ^\prime,z^{\prime i})$ are the
corresponding Fermi coordinates with Baselines $L$ and $L^{^{\prime }}$ then
we have:

$$
\tau^\prime=\tau +\xi ^0(z^\alpha ),\quad z^{\prime j}=z^j+\xi
^j(z^\alpha )
$$
where:
$$
\xi ^0=[1+a_iz^i]^{-1}(\delta \tilde y^0+z^i\delta \Omega _i^0),\quad \xi
^j=-(\delta \tilde y^j+z^i\delta \Omega _i^j),
$$
with:
$$
\delta \tilde y^0\equiv \bar u_\alpha \delta \bar y^\alpha ,\quad \delta
\Omega _i^0\equiv \bar u_\alpha \delta \bar e_i^\alpha ,\quad \delta \tilde
y^j\equiv \bar e_\alpha ^j\delta \bar y^\alpha ,\quad \delta \Omega
_i^j\equiv \bar e_\alpha ^j\delta \bar e_i^\alpha
$$
These transformations are the infinitesimal Born transformations
corresponding to its Fermi realization, $\xi ^\alpha $ being the components
of the appropriate generalization of a Killing vector.

{\em Generalized Killing equations}. A straightforward calculation shows
that the generalized Killing vector satisfies the following generalized
Killing equations:
$$
\nabla _\alpha \xi _\beta +\nabla _\beta \xi _\alpha +{\frac{\partial
g_{\alpha \beta }}{{\partial a_i}}}\zeta _i=0
$$

The Born group defined in this section has a very precise physical meaning,
since it is in fact the special relativistic generalization of the
Irrotational rigid motion group of classical mechanics:

$$
t^\prime=t\quad z^{\prime i}=z^i+\xi ^i(t)
$$
to which it contracts when the speed of light in vacuum is assumed to be infinity.

\section{Generalized Isometries}

Let $g_{\alpha \beta }(x^\rho )$ be the components of a spacetime metric and
let us assume that we can find some functions $\bar g_{\alpha \beta }(x^\rho
,\lambda _a)$, and some functions $f_a(x^\rho )$ belonging to some set $S$
such that we have:

\begin{equation}
\label{Compound}g_{\alpha \beta }(x^\rho )=\bar g_{\alpha \beta }(x^\rho
,f_a(x^\sigma ))
\end{equation}
We shall then say that the $\overline{g}_{\alpha \beta }$ are a Compound
description of the $g_{\alpha \beta }$ with parameter set $S$. Any metric
possesses two limiting Compound descriptions: The {\em Elementary description%
} for which the set $S$ is the empty set and therefore $\bar g_{\alpha \beta
}(x^\rho )=g_{\alpha \beta }(x^\rho )$. And the {\em Trivial description}
for which $\bar g_{\alpha \beta }=g_{\alpha \beta }(f_a)$ with $a=1\cdots 10$%
, and $f_a$ being functions of the four coordinates.

{\em Definition}: If:

\begin{equation}
\label{GeneralizedSymmetry}x^{\prime \alpha }=\varphi ^{\prime \alpha
}(x^\rho ),\quad x^\alpha =\varphi ^\alpha (x^{\prime \rho })
\end{equation}
is a transformation of the spacetime and there exist functions $f_a$ such
that:

$$
g_{\alpha \beta }^\prime(x^{\prime \gamma })=g_{\alpha \beta
}[x^{\prime  \gamma },f_a^\prime (x^{\prime \gamma })]
$$
where:
$$
g_{\alpha \beta }^\prime (x^{\prime \gamma })={\frac{{\partial
\varphi ^\mu }}{{\partial x^{\prime \alpha }}}\frac{{\partial \varphi ^\nu }
}{{\partial x^{\prime \beta }}}}[\varphi ^\delta (x^{\prime \gamma
})]g_{\mu \nu }[\varphi ^\delta (x^{\prime \gamma }),f_a(\varphi ^\rho
(x^{\prime \gamma }))]
$$
then we shall say that the transformations (\ref{GeneralizedSymmetry}) define
a Generalized isometry of the spacetime metric $g_{\alpha \beta }$.

{\em Generalized Killing equations. From} this definition it follows that if:

$$
x^{\prime \rho }=x^\rho +\xi ^\rho (x^\alpha ),\quad f_a^{\prime
}=f_a+\zeta _a(x^\alpha )
$$
are the transformations of an infinitesimal generalized isometry, then the $%
\xi ^\rho $ satisfy a system of differential equations:
$$
\nabla _\alpha \xi _\beta +\nabla _\beta \xi _\alpha +{\frac{\partial
g_{\alpha \beta }}{{\partial f_a}}}\zeta _a=0
$$
which are a generalization of the Killing equations.

Ordinary isometries of an spacetime, when they exist, can be considered as a
limiting case of a generalized isometry. They can be indeed considered as
isometries of the Elementary description. The opposite limiting case is the
group of diffeomorphisms of a 4-dimensional spacetime, since it can be
considered as a Generalized isometry of the Trivial representation of any
metric.

The Born group that we considered in Section \ref{borngroup} is an
intermediate generalized isometry group. The parameter set being in this
case the triads of functions of one single argument.

\section{Generalizations of the Born congruences}

The most obvious generalization of the Born group as it was defined in
Section \ref{borngroup} is to consider any spacetime $V_4$ for which the
Born equations have at least one solution and define the Born group $B(V_4)$
as the group which leaves invariant the metric of $V_4$ written in terms of
the Fermi coordinates with Baseline any worldline of any of the Born
congruences.

We mention below two potentially interesting generalizations of the Born
congruences of any spacetime. Each of these generalizations would lead to a
corresponding generalization of the Born group in the form of a Generalized
isometry of the spacetime on which they are considered. The parameter set of
the corresponding Compound description of the metric depending of course on
the number and type of the arbitrary functions contained in the general
solution of the conditions defining these generalizations.

The first example are {\em Cattaneo's affinities}, which are particularly
interesting from the geometrical point of view. They are defined as those
congruences for which one has:

$$
\hat \nabla _\gamma \Sigma _{\alpha \beta }+b_\gamma \Sigma _{\alpha \beta
}=0,\quad \hat \nabla _\gamma \Sigma _{\alpha \beta }\equiv \widehat{g}%
_\gamma ^\lambda \widehat{g}_\alpha ^\rho \widehat{g}_\beta ^\sigma \nabla
_\lambda \Sigma _{\rho \sigma }.
$$
As it is well known the Born conditions (\ref{borneqs}) express that the
metric of the spacetime $g_{\alpha \beta }$ can be projected onto a metric $%
\hat g_{ij}$ of the quotient manifold $V_3=V_4/{\cal R}$, ${\cal R}$ being the equivalence
relation defined by the corresponding Born congruence. The equations above
express a more general condition. Namely that the linear connection $\Gamma
_{\beta \gamma }^\alpha $ can be projected onto a linear connection of $V_3$%
. This connection is the restriction to these particular congruences of a
more general object first defined by Cattaneo\cite{cattaneo}.

The second example is in our opinion interesting both from the geometrical
and the physical point of view for reasons which will be discussed
elsewhere. We shall say that a timelike congruence is {\em quo-harmonic} if
the tangent vector field $u^\alpha $ is such that there exist three
independent functions $f^i$ solutions of the following equations:

\begin{equation}
\label{adapted}u^\rho \partial _\rho f^i=0,\quad i=1,2,3
\end{equation}

\begin{equation}
\label{SubHarmonic}\hat \triangle f^i-(3-dim)b^\rho \partial _\rho
f^i=0\Leftrightarrow \triangle f^i-(4-dim)b^\rho \partial _\rho f^i=0,
\end{equation}
where:
$$
\triangle \equiv g^{\alpha \beta }\nabla _\alpha \partial _\beta ,\quad \hat
\triangle \equiv \hat g^{\alpha \beta }\nabla _\alpha \partial _\beta
$$
and where $dim=3$. We call these congruences quo-harmonic to distinguish them
from the Harmonic congruences which are defined as those congruences for
which there exist three independent functions $f^i$ solutions of equations
(\ref{adapted}) and (\ref{SubHarmonic}) with $dim=4$. Harmonic congruences were
extensively discussed in ref.\cite{belcoll} where it was proved in
particular that irrotational Born congruences are never harmonic. This is
the reason why harmonic congruences do not qualify as appropriate
generalizations of the Born congruences. On the contrary, every Born
congruence is quo-harmonic. Indeed, using any system of adapted coordinates,
i.e. any system of coordinates for which $u^i=0$, eqs. (\ref{adapted}) and (\ref
{SubHarmonic}) become:

$$
\partial _0f^i=0,\quad \hat \triangle f^i=0,
$$
where $\hat \triangle $ is here the Laplacian of the quotient metric $\hat
g_{ij}(x^k)$. The above equations always have local triads of independent
solutions and this proves our assertion.


\begin{thebibliography}{99}

\bibitem{Bel1} Ll.\ Bel in {\it Relativity in General}, Eds. J.\ Diaz, M.\ Lorente, Editions
    Fronti\`{e}res, 47 (1994)

\bibitem{killing}  Killing, W., Journal f\"ur die reine und angew.
Math.(Crelle), {\bf 109}, pp. 121-186 (1892)

\bibitem{born}  Born, M., Ann. Physik (4), {\bf 30}, p. 1 (1909)

\bibitem{herglotz}  Herglotz, G.,Ann. Physik (4), {\bf 31}, p. 393 (1910)

\bibitem{noether}  Noether, F.,Ann. Physik (4), {\bf 31}, p. 919 (1910)

\bibitem{fermi}  Fermi, E., R.C. Acc. Naz. Lincei, {\bf 31}, 21-23, 51-52
(1922)

\bibitem{walker}  Walker, A.G., Proc. Roy. Soc. Edinb., {\bf 52}, p. 345
(1932)

\bibitem{mast}  Mast, C.B. and Strathdee, J., Proc. Royal Soc., A, {\bf 252}%
, pp.476-487 (1959)

\bibitem{cattaneo}  Cattaneo, C., Ann. di Mat. pura ed appl., S, IV, T. {\bf %
XLVIII}, p. 361 (1959)

\bibitem{synge2}  Synge J.L., Relativity: {\em The General Theory}, North
Holland (1960). And references therein.

\bibitem{moller}  M\o ller C., {\em The theory of Relativity}, Oxford
University Press (1962)

\bibitem{manasse}  Manasse, F.K. and Misner, C.W., J. Math. Phys., {\bf 4},
p. 735 (1963)

\bibitem{synge}  Synge J.L., Relativity: {\em The Special Theory}, North
Holland (1965). And references therein.

\bibitem{Ni}  Ni, W.T. and Zimmermann, M., Phys. Rev. D, {\bf 17}, p. 1473
(1978)

\bibitem{belcoll}  Bel, Ll. and Coll, B., Gen. Rel. Grav., {\bf 26}, 6, p.
613 (1993)
\end{thebibliography}
\end{document}